\documentclass[12pt]{revtex4-1}

\usepackage{graphicx}
\usepackage{dcolumn}
\usepackage{bm}
\usepackage{pstricks,epsfig}
\usepackage{varioref}
\usepackage{amssymb}
\usepackage{amsthm}

\begin{document}

\title{Electronic structure of gadolinium complexes in ZnO in the GW approximation}

\author{A. L. Rosa}
\affiliation{Federal University of Goi\'as, Institute of Physics, Campus Samambaia, 74690-900, Goi\^ania, Goi\'as, Brazil}
\affiliation{BCCMS, Universit\"at Bremen, Am Fallturm 1, 28359, Bremen, Germany}
\author{Th. Frauenheim}
\affiliation{BCCMS, Universit\"at Bremen, Am Fallturm 1, 28359, Bremen, Germany}

\begin{abstract}

The role of intrinsic defects has been investigated to determine
binding energies and the electronic structure of Gd complexes in
ZnO. We use density-functional theory and the GW method to show that the presence of vacancies and 
interstitials affect the electronic structure of Gd doped ZnO. However, the strong
localization of the Gd-$f$ and $d$ states suggest that carrier
mediated ferromagnetism in this material may be difficult to achieve.

\end{abstract}
\maketitle

\section{Introduction}

Doping ZnO with rare-earth elements has been widely used to tailor its
electronic, magnetic and optical properties. In particular, the long
life times of the excited states allow for an easy realization of
population inversion with promising applications in optoelectronics
\cite{Ronning:10} as well as spintronics\,\cite{Dietl2000}. Channeling
experiments as well as photoluminescence studies
\cite{Wahl:03,Du:08,Ji:09,Ebisawa:08,Ronning:10,Petersen:10,Dai:11,Tsuji:12,Mezdrogina:12}
indicate, that rare-earth elements in ZnO are preferentially
incorporated at cation sites. The incorporation of Gd in the presence
of intrinsic defects in ZnO could lead to the formation of stable
compounds inside the
crystal\,\cite{Ney:12,Bantounas:11,Bantounas:2014,Zhang:14,Flemban:16,Roqan:15}.
In particular, the formation of such complexes involving gadolinium
and oxygen vacancies or zinc interstitials could promote
ferromagnetism in ZnO as suggested in
Refs.\,\cite{Bantounas:2014,Roqan:15}. This idea has been reinforced
by investigations of Gd doped ZnO thin films grown by pulsed laser
deposition which suggested that the formation of Gd-O vacancies
complexes may be formed under oxygen deficient
conditions\,\cite{Flemban:16,Roqan:15}. However, magnetic properties
of ZnO doped with Gd have been investigated using ion implantation
\,\cite{Potzger2006} and pulsed-laser deposition
(PLD)\,\cite{Ungurearnu2007} and the samples were found to be
paramagnetic. X-ray spectroscopy (XPS) and Fourier Transform Infra Red
(FT-IR) spectroscopy suggested that nearest Gd ions do not take part
in carrier mediated ferromagnetism\cite{Ravi2016}. On the theoretical
side, DFT calculations using the generalized-gradient approximation
(GGA) and GGA plus Hubbard U\,\cite{Ney:12} show no coupling between
Gd atoms rendering Gd doped ZnO to be paramagnetic in the absence of
defects. Under O-deficient conditions Roqan {\it et
  al.}\cite{Roqan:15} used GGA plus U and hybrid functionals to show
that either oxygen vacancies or zinc interstitials could stabilized
ferromagnetism in Gd doped ZnO. Therefore, a more clear understanding
of formation energies and electronic structure of Gd complexes in ZnO
is needed to clarify these aspects.

In this letter, we investigate the role of oxygen and zinc vacancies
and interstitials using DFT and the GW method in order to determine changes
in the magnetic properties and electronic structure of Gd
doped ZnO. We show that although there is a strong localization of the
Gd states with no significant change in its magnetic moment, other
than in the presence of oxygen interstitial defects. Furthermore, the
electronic struture shows negligible overlap of Gd states with the ZnO
matrix, suggesting that carrier mediated ferromagnetism in this
material may be difficult to achieve.

\section{Computational details}

We have used density-functional theory (DFT)\cite{Kohn:65}
together with the projected augmented wave (PAW) method, as
implemented in the Vienna Ab initio Simulation Package (VASP)
\cite{Kresse:99}. The Perdew-Burke-Ernzerhof (PBE) form of the
generalized gradient approximation (GGA) for the exchange-correlation
potential was used to obtain geometries, formation energies and
magnetic moments. To model Gd impurities and Gd complexes in ZnO we
built up a 72 atom supercell using our calculated PBE lattice
parameters of ZnO, $a$=3.25{\AA} and $c$=5.25{\AA}. To ensure
convergence of structural, electronic and magnetic properties, a
cutoff of 400 eV was used for the plane-wave expansion of the wave
function. The criteria on force convergence was 0.01\,eV/{\AA}. For
Brillouin zone integrations, a $(3\times 3 \times 3)$ Monkhorst-Pack
k-point sampling was used.  The Gd-5$s$, -5$p$, and -4$f$
electrons are treated as a valence shell, as well as the Zn-3$d$
electrons.

For the determination of the electronic structure we have used the GW
method\,\cite{Shishkin:06}. The wave functions are kept fixed to the GGA
level, whereas the eigenvalues are updated in the Green's function
only.  A cutoff of 200 eV for the response functions, as well as 1024
bands have been employed. For ZnO in particular, we obtain a band gap
of 3.3\,eV and Zn-3d states at 7.0\,eV which is in reasonable agreement
with the experimental value of 3.44\,eV \cite{madelung:parameters}, as
well as with other all-electron GW0
calculations\,\cite{Friedrich:11,Shih:10,Lany:10,Shishkin:07}

\section{Results}

\begin{figure}[ht!]
\pspicture(0,0)(16,7)
\rput[bl](0,4){\epsfig{file=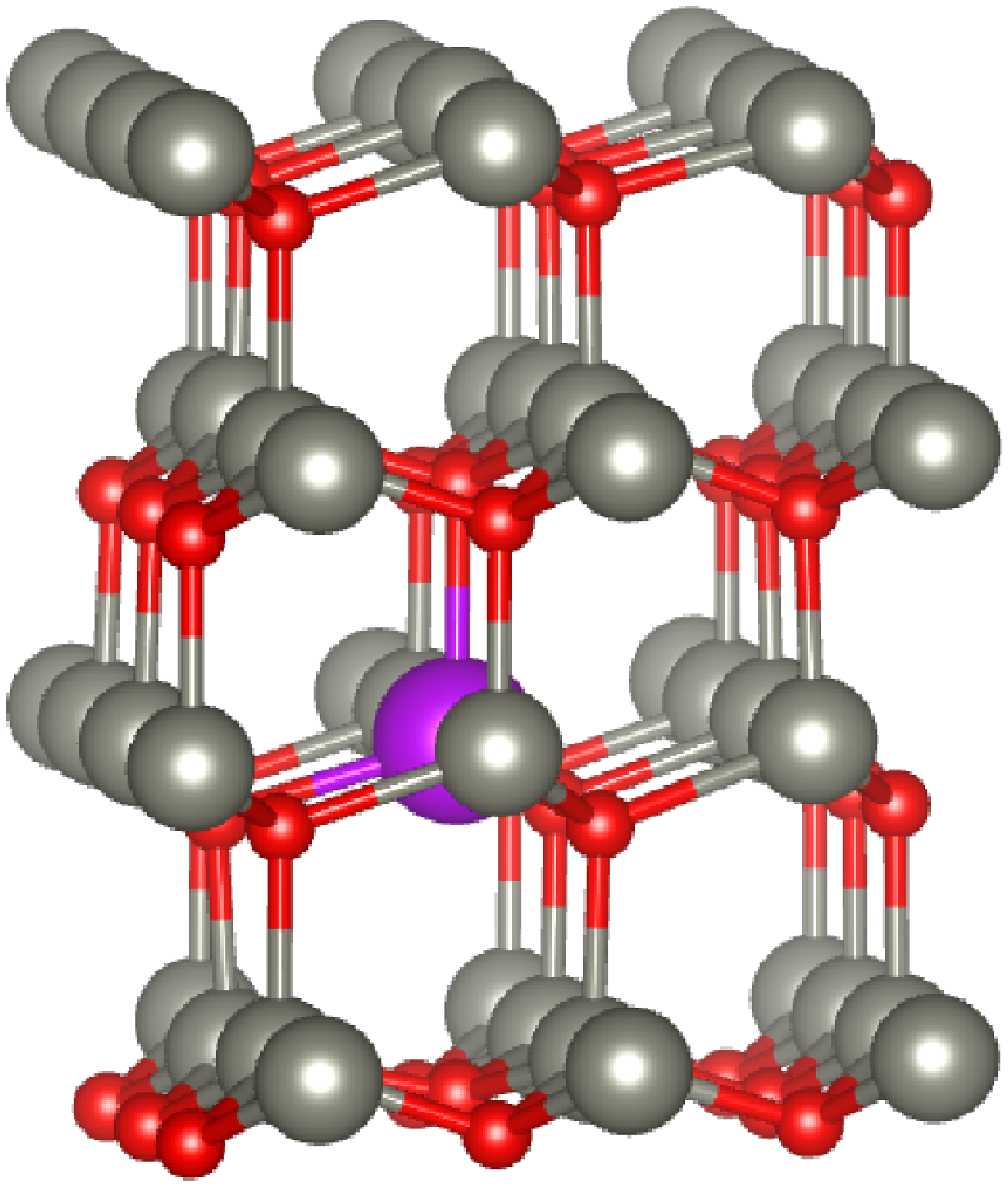,width=2.5cm}}
\rput[bl](3,4){\epsfig{file=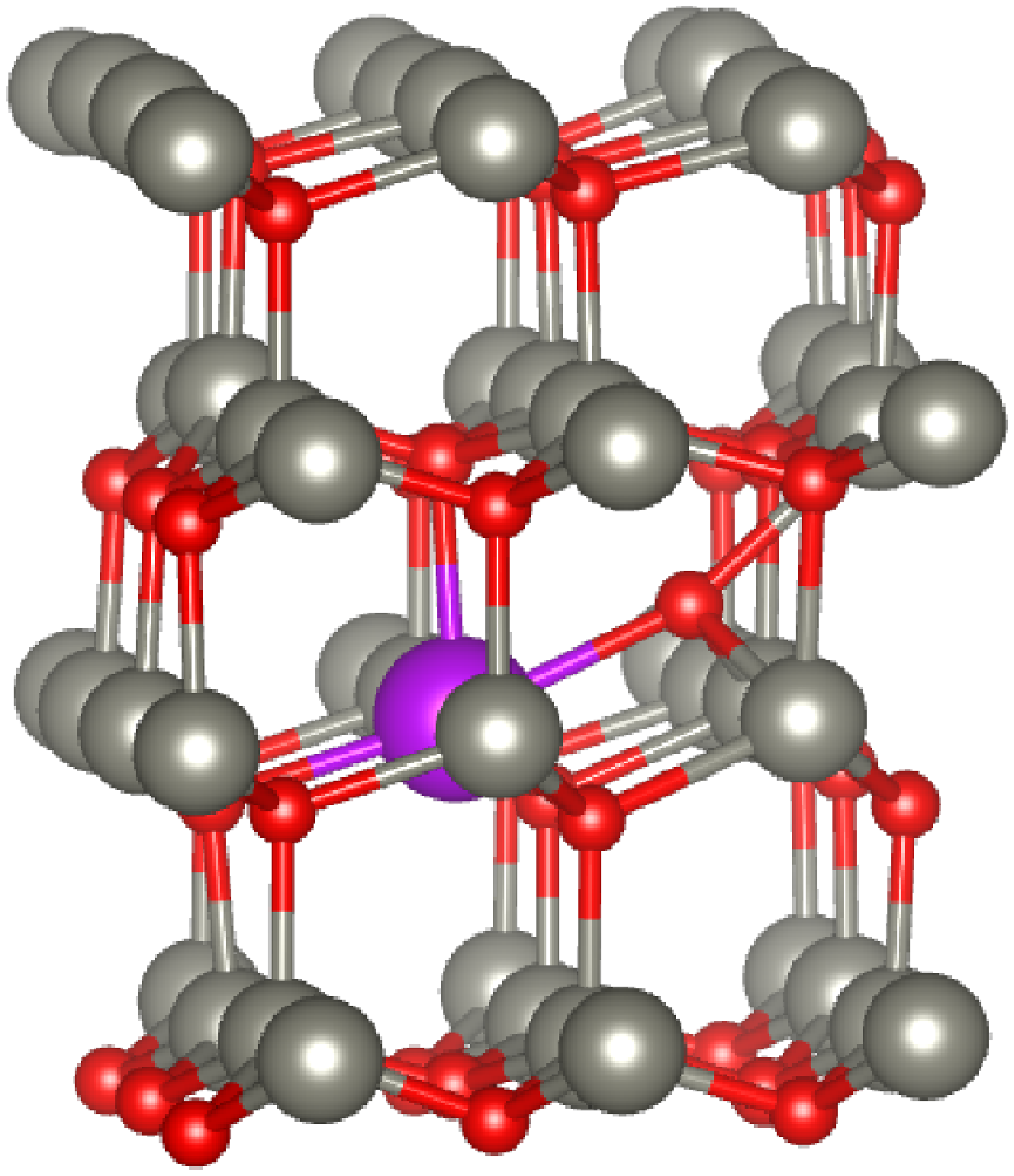,width=2.5cm}}
\rput[bl](6,4){\epsfig{file=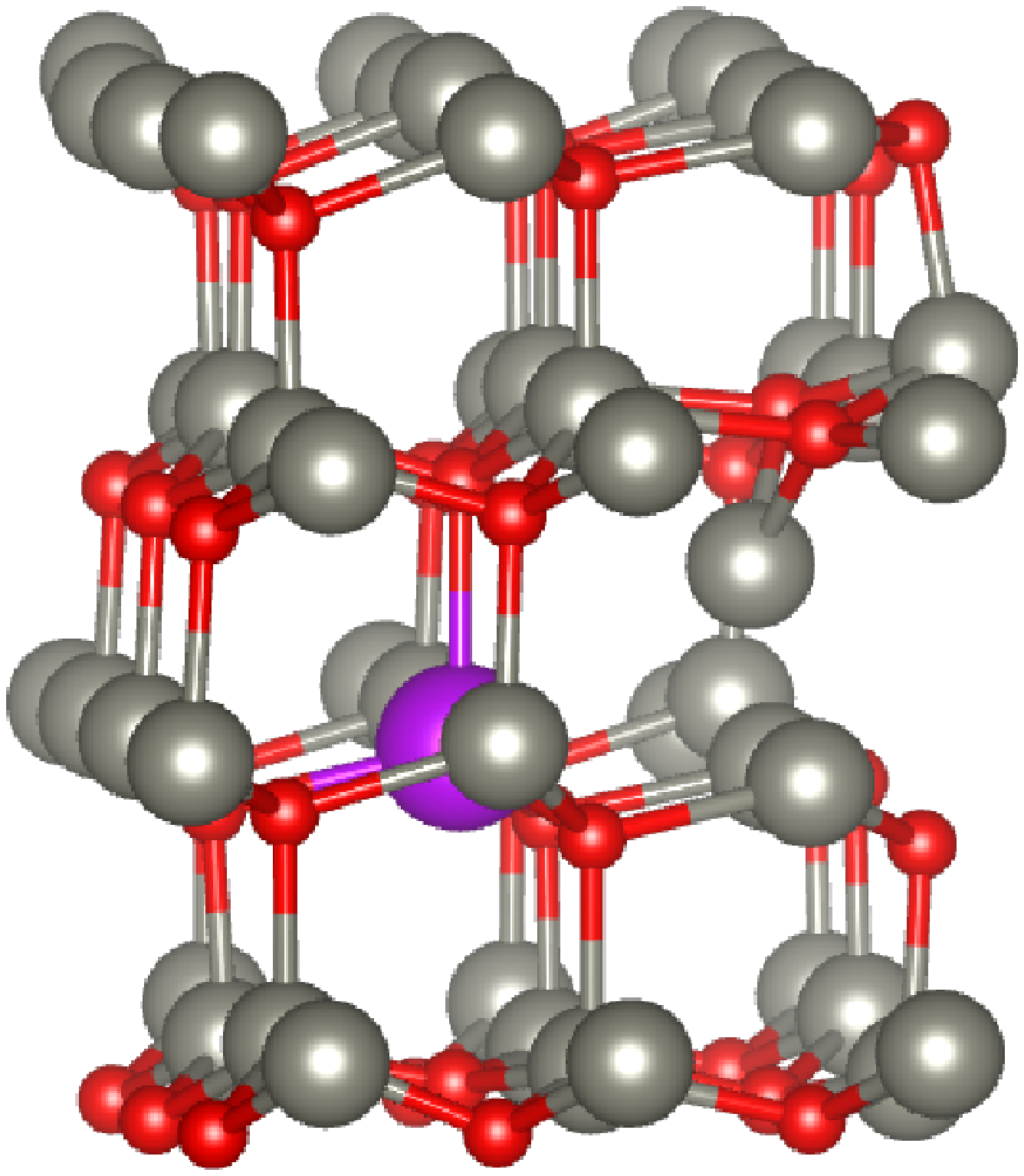,width=2.5cm}}
\rput[bl](9,4){\epsfig{file=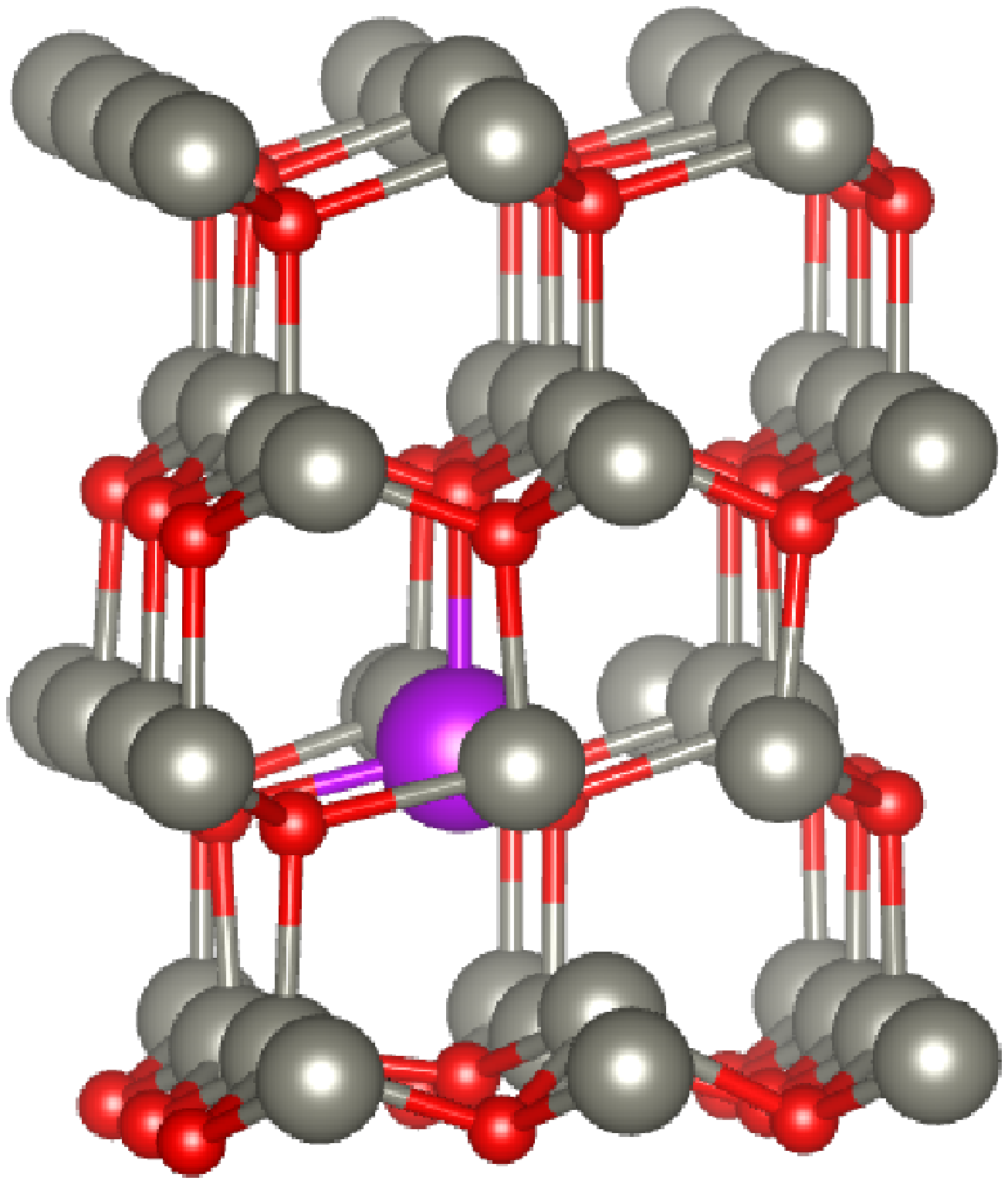,width=2.5cm}}
\rput[bl](12,4){\epsfig{file=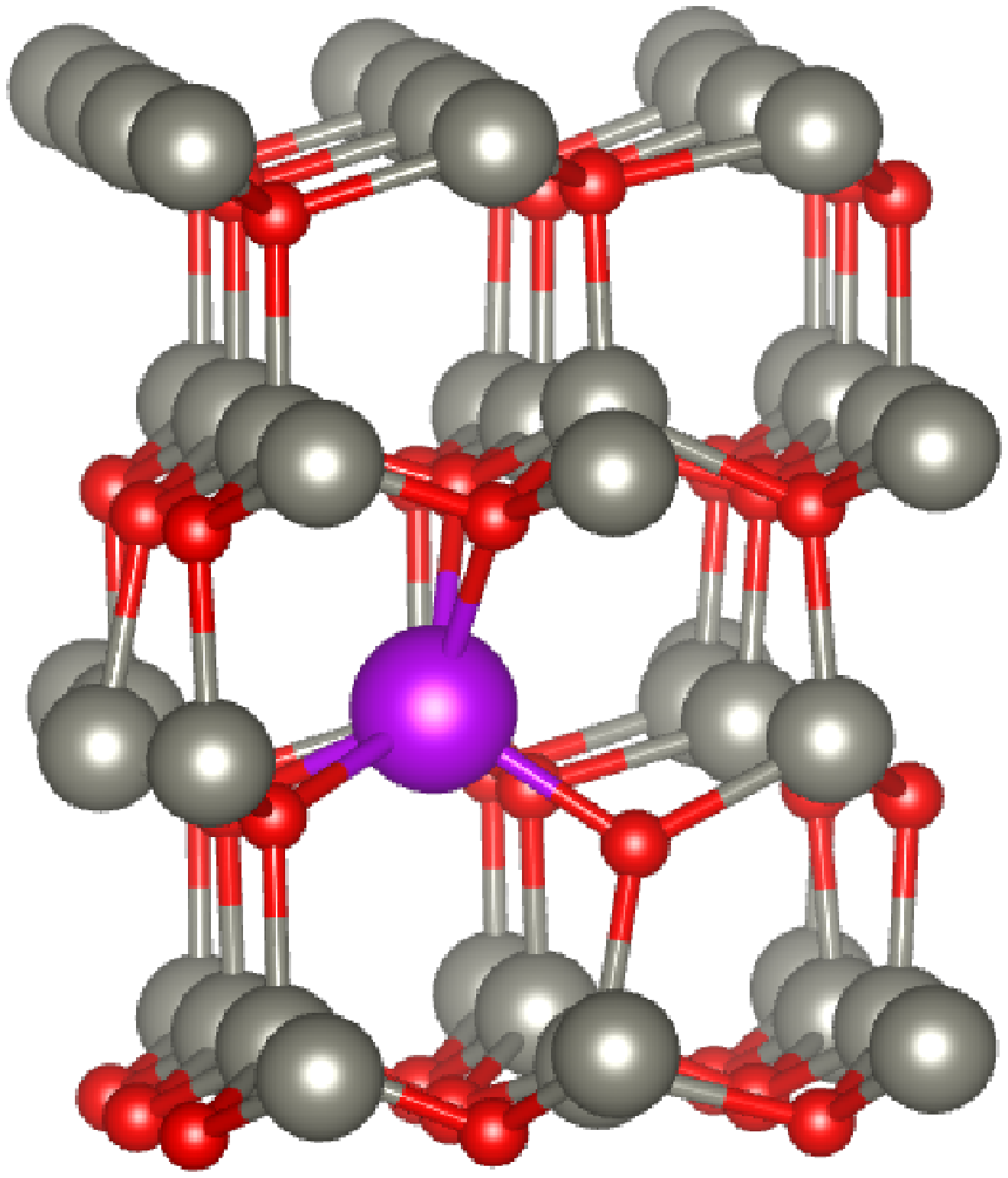,width=2.5cm}}
\rput[bl](0,0){\epsfig{file=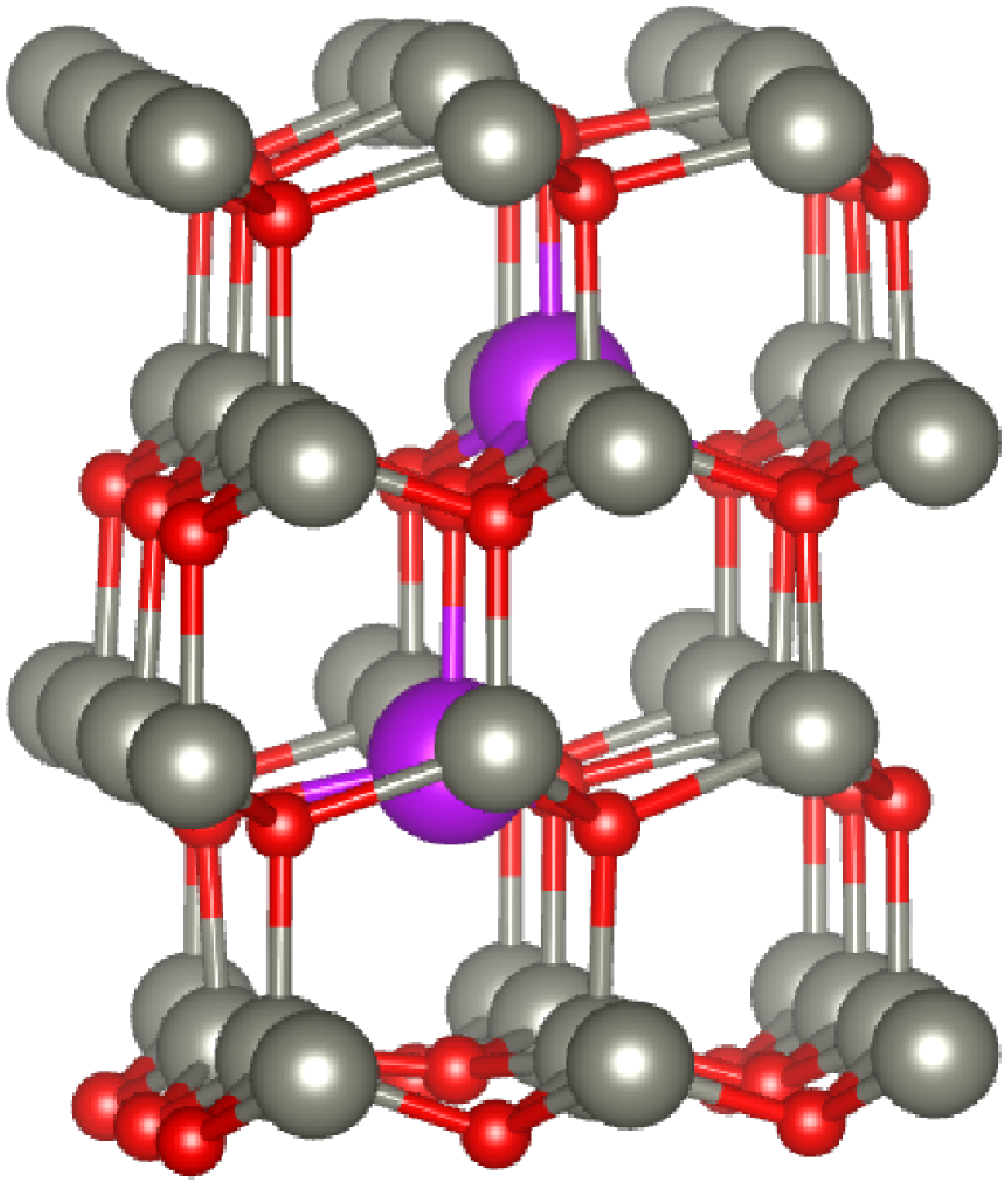,width=2.5cm}}
\rput[bl](3,0){\epsfig{file=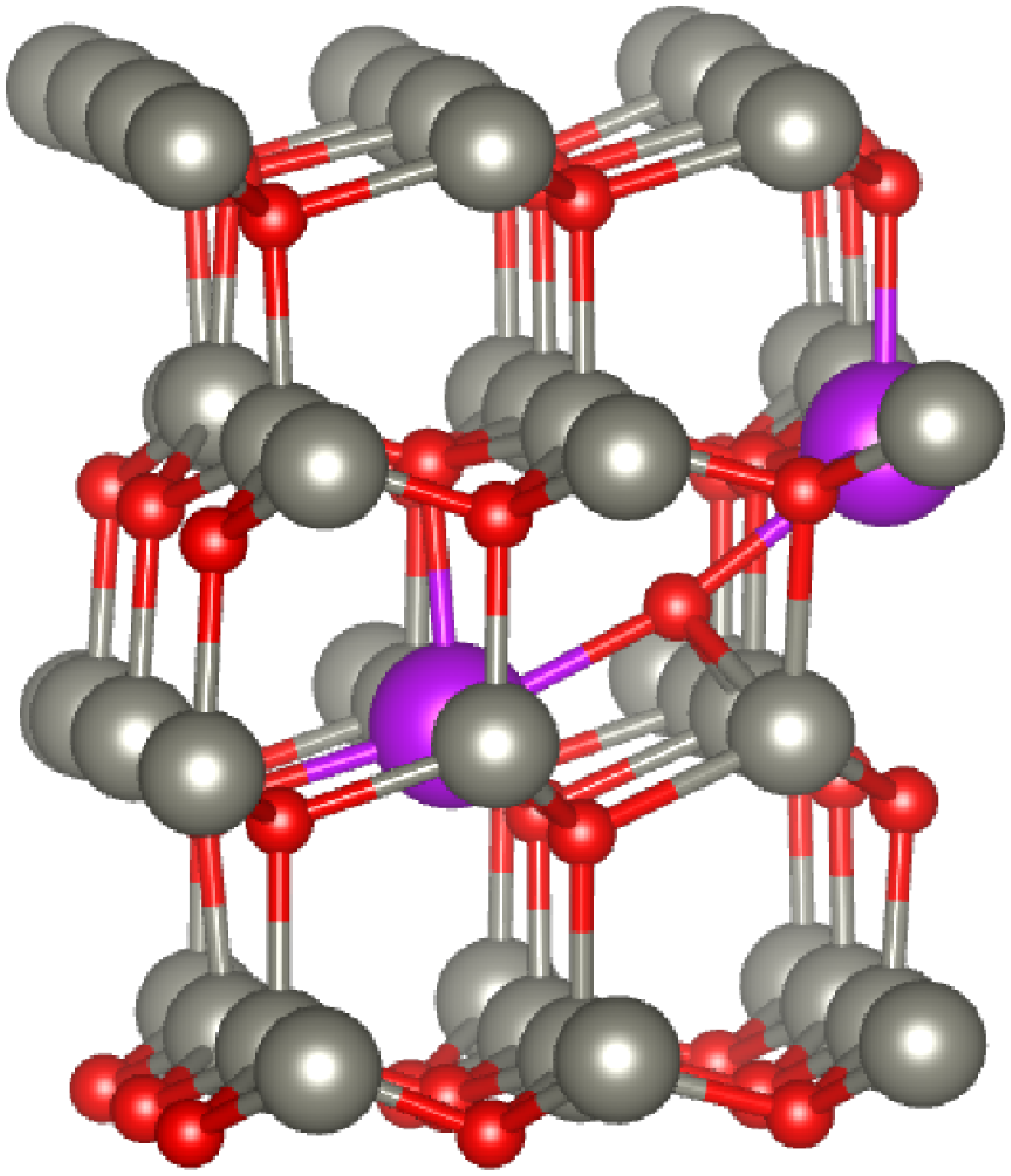,width=2.5cm}}
\rput[bl](6,0){\epsfig{file=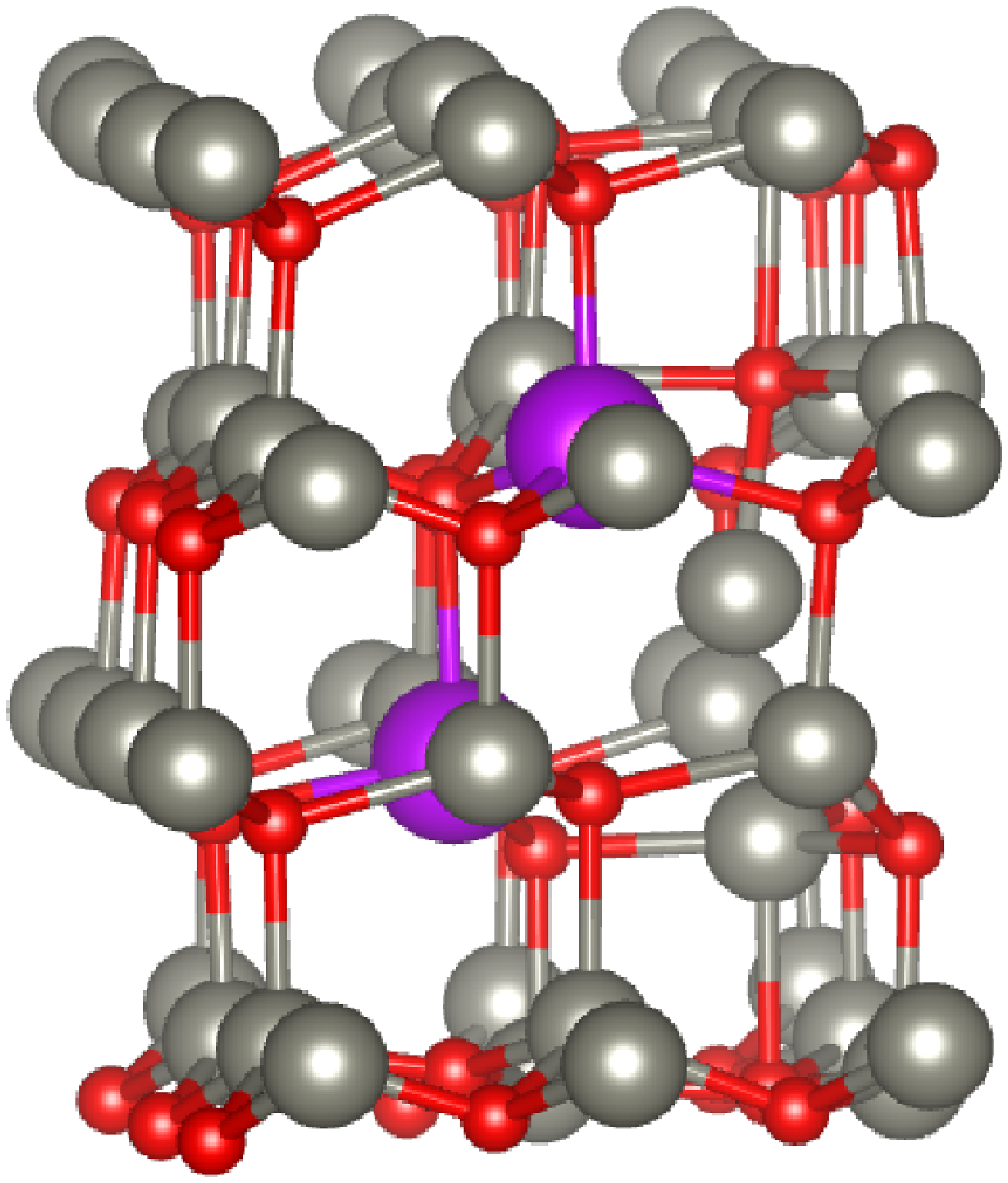,width=2.5cm}}
\rput[bl](9,0){\epsfig{file=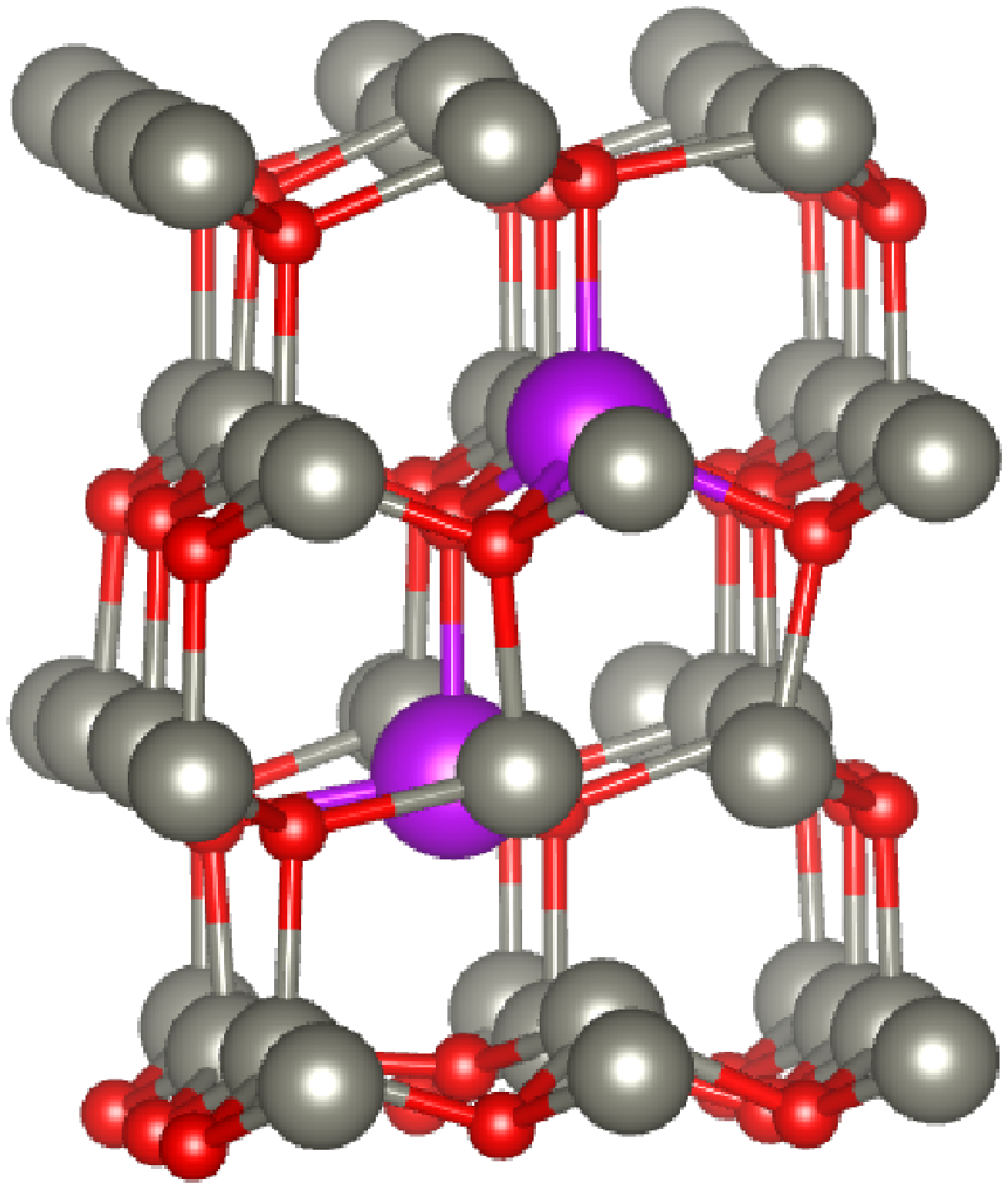,width=2.5cm}}
\rput[bl](12,0){\epsfig{file=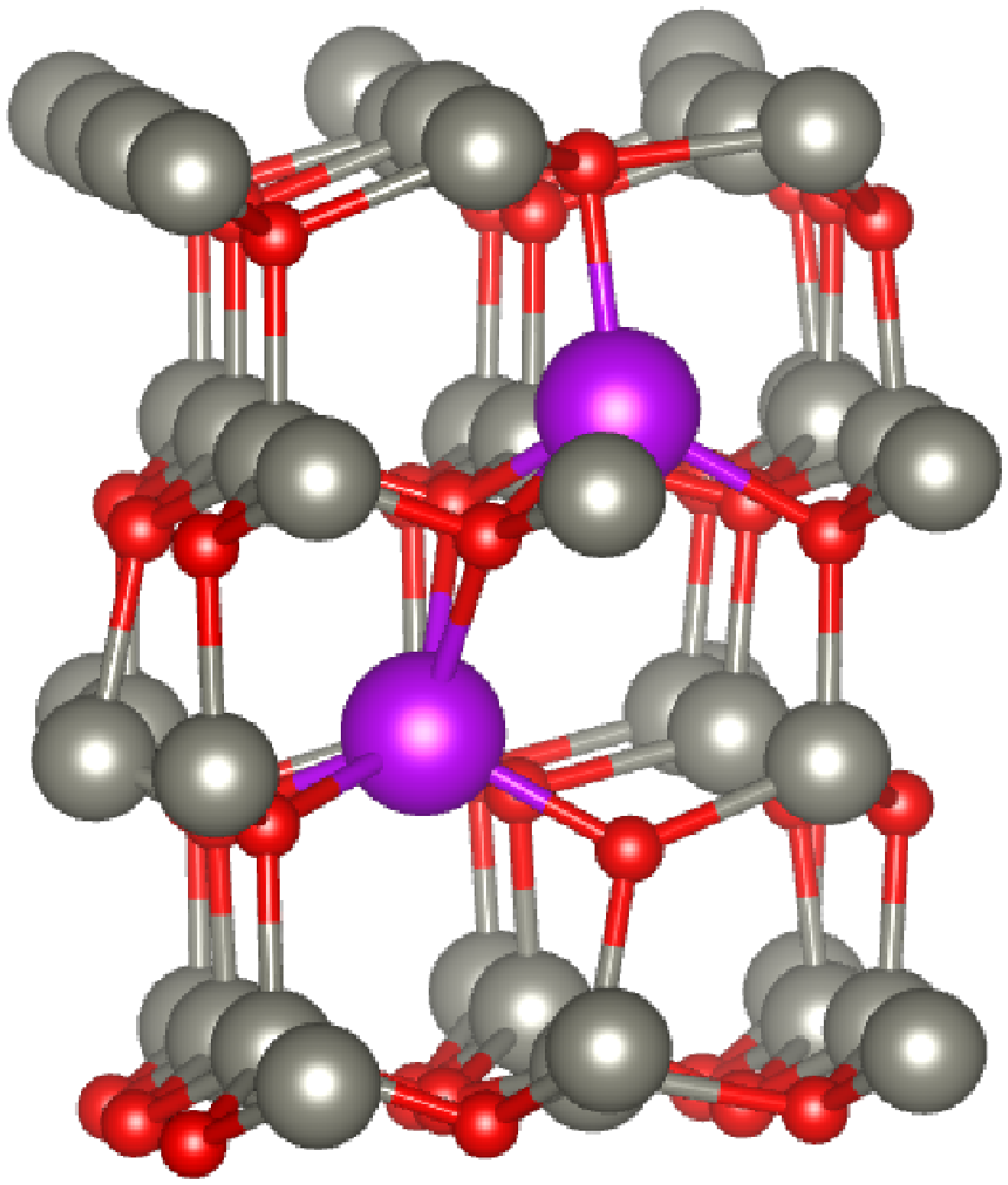,width=2.5cm}}
\rput[bl](0,7){a)}
\rput[bl](3,7){b)}
\rput[bl](6,7){c)}
\rput[bl](9,7){d)}
\rput[bl](12,7){e)}
\rput[bl](0,3){f)}
\rput[bl](3,3){g)}
\rput[bl](6,3){h)}
\rput[bl](9,3){i)}
\rput[bl](12,3){j)}
\endpspicture
\caption{\label{fig:geometries} Relaxed geometries within PBE for Gd complexes in ZnO. Gray are Zn, red are O and magenta are Gd. a) ${\rm Gd_{Zn}}$, b) ${\rm Gd_{Zn} + O_i}$, c) ${\rm Gd_{Zn} + Zn_i}$, d) ${\rm Gd_{Zn} + V_{O}}$ e) ${\rm Gd_{Zn} + V_{Zn}}$, f)${\rm 2Gd_{Zn}}$, g) ${\rm 2Gd_{Zn} + O_i}$, h)${\rm 2Gd_{Zn} + Zn_i}$, i) ${\rm 2Gd_{Zn} + V_{O}}$, j) ${\rm 2Gd_{Zn} + V_{Zn}}$}.
\end{figure}

The geometry of the structures we have investigated is shown in
Figs.\,\ref{fig:geometries}. From Fig. \ref{fig:geometries}(a) we see
that a single Gd occupying a substitutional Zn sites does not produce
strong distortion in the ZnO lattice. The Gd-O bond lengths remain
very close to the values in pure ZnO. We obtain 2.17-2.24\,{\AA}.  If
a second Gd atom is added at a near Zn site, as shown in Fig. \ref{fig:geometries}(f) a
negligible relaxation is seen. The Gd-O bond lengths are in the range
2.15-2.21\,{\AA}. This is due to the strong
localization of the Gd-$f$ states with negligible overlap with the ZnO
lattice other than the oxygen nearest neighbors.

An oxygen interstitial atom at the octahedral position near a Gd atom as shown in
Fig.\,\ref{fig:geometries}(b) increases Gd coordination from four-fold
to five fold.  The bond Gd-O lengths are in the range 2.18-2.23\,{\AA}. The
ZnO cage where the oxygen atom is inserted relaxes inwards to
accomodate the Gd-O bonds.  Adding again a Gd atom along the
ZnO $c$ axis to form a complex containing two Gd atoms and
an O interstitial (${\rm 2Gd_{\rm Zn}+O\_i}$), as shown in
Fig. \ref{fig:geometries}(g) the Gd-Gd distance is 4.41\,{\AA}. The Gd-O distances lies between 2.19 and
  2.25\,{\AA}. As a
matter of comparison in Gd$_2$O$_3$ this distance is
2.39\cite{Nelson}. There is a distortion of the oxygen at a octahedral site,
  which now is bonded to two Gd atoms.  Due to the strain around the
  Gd-O-Gd complex, some Zn atoms relax outwards the defect complex. We
  should point out that this is a different defect from the one with
  two gadolinium atoms stacked along the $c$ direction separated by an
  oxygen atom described in Refs.\,\cite{Bantounas:2014}.

A complex involving a Gd substitutional and a Zn interstitial atom
(${\rm Gd_{\rm Zn}+Zn_i}$), shown in Fig.\,\ref{fig:geometries}(c),
leads to a huge distortion of the Zn interstital atoms and also other
Zn atoms around the Gd substitutional, although Gd itself is not much
displaced. Gd-O bond lengths are found to be 2.15-2.25{\AA}. A
configuration where two Gd atoms sit close to a Zn interstitial atoms
(${\rm 2Gd_{\rm Zn}+Zn_i}$), shown in Fig.\,\ref{fig:geometries}(h),
leads to a huge distortion in the lattice. The Zn atom moves away from
the Gd atoms and from the defect. The Gd-Gd distance is 3.57\,{\AA}. 

Next we consider the removal of an oxygen atom near a Gd atom as shown
in Fig.\,\ref{fig:geometries}(d) forming a ${\rm Gd_{Zn}+V_O}$
complex.  There is a clear relaxation of the surrounding zinc atoms
towards the oxygen vacancy, as expected. The Gd substitutional atom
shows a very small relaxation. This configuration is not expected to
be favorable, as we will discuss later, since it decreases the Gd
coordination number. Gd-O bond lengths vary between 2.12 and
2.16\,{\AA}.  By adding an extra Gd atom in order to have a ${\rm 2Gd_{Zn}+V_O}$ complex
(Fig.\,\ref{fig:geometries}(i)) allows the Gd atoms to move further apart and
a slight relaxation of the oxygen atoms towards the Gd is seen. Gd-O
bond lengths are 1.99 and 2.30{\AA}. 

Finally we consider the formation of a complex with a Gd
substitutional and a Zn vacancy  ${\rm Gd_{Zn}+V_{Zn}}$ as has been suggested
in Ref.\,\cite{Bantounas:2014}. The Gd-O distances lie between 2.10 and
2.30\,{\AA}. The structure is somewhat distorted because the Gd atom
which is initially substitutional, moves towards the center of the
cage to increase their coordination with nearby oxygen atoms, as shown in
Fig.\,\ref{fig:geometries}(e). Adding a Gd atom at a nearby Zn site leads to a distortion of the both oxygen and gadolinium atoms to better
accomodate the bond angles. Gd-O atoms lie between 2.15-2.43\,{\AA}

To verify the thermodynamic stability of the
investigated defect complexes, we follow the approach derived by van
de Walle and Neugebauer\cite{JAP:2004}.  The binding energy is calculated according to ${\rm E_b =  E_{f}^{complex} - \sum_i{E_{f}^{defect}}}$ where ${\rm E_{f}^{complex}}$ is the formation energy of a defect complex and ${\rm E_{f}^{defect}}$ is the formation energy of an isolated defect (Gd impurity or intrinsic defect).

The incorporation of a Gd atom at a Zn site costs 2.61 (9.86) eV under
O-rich (O-poor) conditions. The Gd magnetic moment remains almost
unchanged and has a value of 6.92 $\mu_{\rm B})$. The addition of
another Gd atom at a Zn second nearest neighbor is 1.34 eV higher in
energy. The total magnetic moment of 14 $\mu_{\rm B})$ (7 $\mu_{\rm
  B})$/Gd) reflects the negligible interaction between Gd atoms in the
cell. Table\,\ref{table:formation} shows the binding energy of Gd in
ZnO under the presence of intrinsic defects. One can identify some
defect complexes with low formation energy under O-poor conditions, the ${\rm Gd_{Zn} + V_{Zn}}$ and under O-rich conditions the  ${\rm Gd_{Zn} + O_i}$.
The ${\rm Gd_{Zn} + O_i}$ complex is the only defect
able to change the Gd magnetic moment due to the overlap of Gd with
its nearest neighbor oxygen atoms. The magnetic moment on Gd is 6.02
${\rm \mu_B}$ and 0.72 ${\rm \mu_B}$ on the $p$ orbitals of
oxygens. Around 0.1 ${\rm \mu_B}$ is distributed on the 5$d$ orbitals
of Gd.  Murmu {\it et al},\cite{Murmu:12} suggested that after
annealing a reduction of Gd can be observed and attributed it to Gd
clustering. Here we suggest that the formation of a
${\rm Gd_{Zn} +  O_i}$ complex can be reponsible for this reduction. We find a
similar behavior for Eu doping in ZnO\,\cite{Lorke:16,Ronning2014}.
Also, Roqan\cite{Roqan:15} suggested that oxygen deficient defects may
be responsible for mediating ferromagnetism in ZnO.  Our results
indeed show that this defect has a low formation energy in
ZnO. However, the there is no direct exchange interaction between
Gd-$f$ electrons, so they cannot be responsible for take part in
carrier mediated ferromagnetism. We have further considered the
complexes in the -1 and +1 and 2+ charge states (to be published).
These defects are not energetically favorable neither under O-rich and
nor Zn-rich conditions and have no influence on the Gd magnetic moment
atoms, indicating strong localization of $f$ orbitals.

\begin{table}[ht!]
\begin{center}
\caption{\label{table:formation} \bf Total magnetic moments  $\mu_{\rm tot}$ (in $\mu_{\rm B})$ and binding energies $\rm E_b$ (in eV) of neutral Gd complexes in ZnO calculated at PBE level under O-rich and  O-poor conditions, respectively.}
\begin{tabular}{lccc}
\hline
\hline
complex   & $\mu_{\rm tot}$ & $\rm E_b$\\
\hline
${\rm Gd_{Zn} + O_i}$ &    6.02 &  -1.65   \\
${\rm Gd_{Zn} + Zn_{i}}$ &  6.86  & 2.05  \\
${\rm Gd_{Zn} + V_{O}}$ &   6.99 & 1.13\\
${\rm Gd_{Zn} + V_{Zn}}$ &  6.98 &  -0.97\\
${\rm 2~Gd_{Zn} + O_i}$       &   14.00   & -4.43 \\
${\rm 2~Gd_{Zn} + Zn_i}$       &    14.00   &  1.26   \\
${\rm 2~Gd_{Zn} + V_{O}}$       &  14.00   &   1.04    \\
${\rm 2~Gd_{Zn} + V_{Zn}}$       &    14.00   &   -2.92    \\
\hline
\hline          
\end{tabular}
\end{center}
\end{table}

\begin{figure}[ht!]
\begin{center}
\pspicture(0,0)(16,12)
\rput[bl](0,0){\epsfig{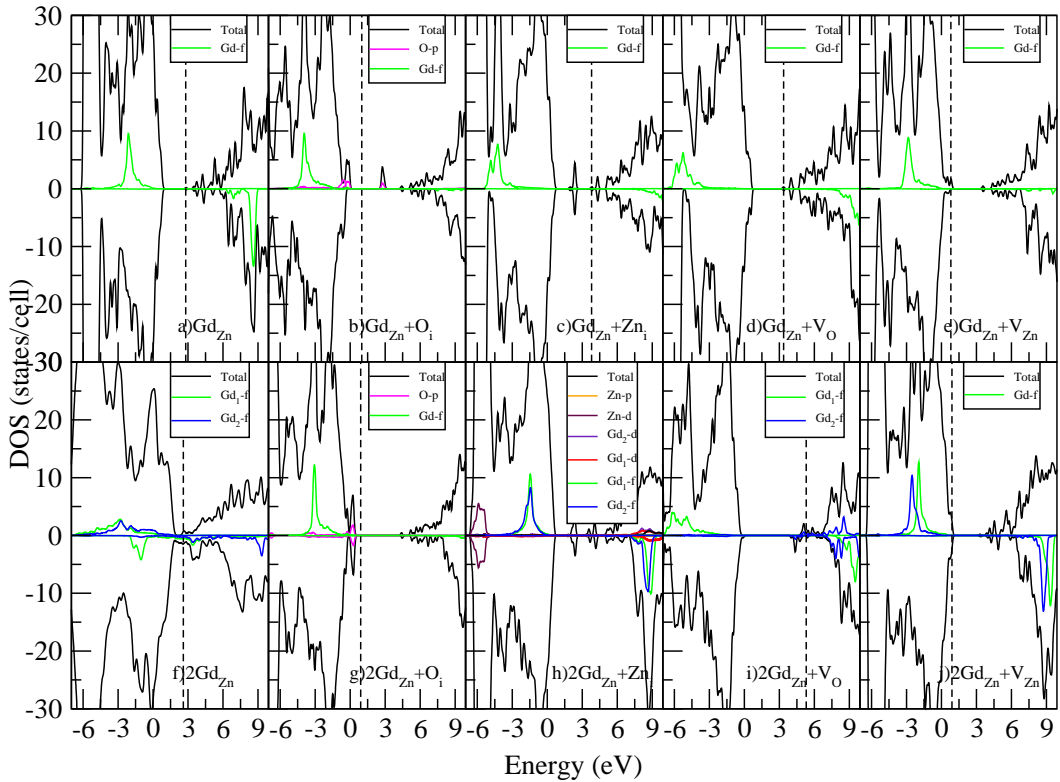}}
\endpspicture
\caption{\label{fig:dos} Density of states for (a) substitutional Gd atom in ZnO, (b) substitutional Gd plus an oxygen interstitial atom, (c) substitutional Gd plus a zinc interstitial atom, (d)  substitutional Gd plus an oxygen vacancy, (e)  substitutional Gd plus a zinc vacancy, (f)  two substitutional Gd atoms in ZnO, (g) 2 substitutional Gd atoms plus an oxygen interstitial atom, (h) two substitutional Gd atoms plus a zinc interstitial atom, (i)  two substitutional Gd atoms plus an oxygen vacancy and (j)  two substitutional Gd atoms plus a zinc vacancy. The vertical dashed line represents the Fermi level. }
\end{center}
\end{figure}

Next we discuss the electronic structure of the above mentioned
neutral complexes. In Fig.\,\ref{fig:dos}(a)-(j) the density of states (DOS)
of Gd-doped ZnO calculated with GW@PBE for Gd concentrations of 3.7\%
and 5.6\,\%.  Due to the strong localization of the Gd-4$f$ orbitals, the
degree of mixing usually is small. Therefore, any
perturbation to the crystal due to the presence of intrinsic defects
should overlap with these states in order to delocalize the
Gd-$f$states.

The presence of a single Gd atom at a zinc site, shown in
Fig.\,\ref{fig:dos}(a) leads to no change in the Gd-f states. The $f$
spin-up states are fully occupied, giving a total magnetic moment of
6.9${\rm \mu_B}$. Furthermore, the Gd spin splitting is around
10\,eV.  The presence of two Gd atoms as shown in
Fig.\,\ref{fig:dos}(f) leads to completely filled Gd-$f$ up states but
partially filled $f$-down states. However, the magnetic moment of Gd
does not change, because there is little overlap neither between the
Gd atoms nor between the Gd and the ZnO crystal. 

Next we show the electronic structure for the structure ${\rm Gd_{Zn}  + O_i}$.  The band structure of this defect is shown in
Fig. \,\ref{fig:dos}(b). We find the occupied Gd-$f$ spin up states
are still located within the valence band of ZnO, while the Gd-$f$
down states lie well above the ZnO conduction band. The presence of an
additional oxygen atom nearby a Gd can promote a change in the
oxidation state of Gd by populating the ${\rm O_{int}-p}$ orbitals,
which lie right above the top of the valence band.  Similarly, we have
recently show that the presence of a nearby oxygen atom can modify the
europium spin magnetic moment in Eu doped ZnO\,\cite{Lorke:16}.
As an additional Gd atom is added to this complex to form the
${\rm 2Gd_{Zn} + O_i}$, the electronic structure changes. The O-2$p$ state present in the band
gap moves towards the ZnO valence band below the Fermi level, as shown
in Fig.\,\ref{fig:dos}(g).

For the ${\rm Gd_{Zn} + Zn_i}$ complex new states are introduced in
the band gap due to presence of the defect, as shown in
Fig.\,\ref{fig:dos}(c). However, no overlap between Gd states in
found.  The incorporation of an extra Gd atoms to form the ${\rm
  2~Gd_{Zn} + Zn_i}$ complex produces little change in the band
structure. This defect has been suggested in Ref.\cite{Roqan:15} to
have overlap between Zn-$p$ and -$d$ states and Gd-$f$ states in the
conduction band. Our GW@PBE results show that this overlap is
negligible as it can be seen in Fig.\,\ref{fig:dos}(h).  Altough the
splitting between the Gd-$f$ spin up and spin down is significantly
reduced to about 3.6\,eV, the Zn-$d$ states lie well below the Gd
states deep in the valence band an does not overlap with the Gd-f
states, as found in Ref. \cite{Roqan:15}. A possible explanation for
this discrepancy is that the description of Zn-$d$ states is well
known to have a dependence on the choice of U\,\cite{Janotti:09,Janotti:07}, which may affect its relative position with respect to the valence band top.

Next we consider the ${\rm Gd_{Zn} + V_{O}}$ complex. The electronic
structure is similar to the one without the oxygen vacancy with a
large Gd $f-f$ spin splitting of 10\,eV (see
Fig.\,\ref{fig:dos}(d). The oxygen vacancy states appear 2.6 eV above
the top of the valence band. The introduction of a second Gd atom to produce the
 ${\rm 2Gd_{Zn} + V_{O}}$ complex
nearby disturbs the ZnO lattice and shifts the Fermi level
to within the conduction band, as it can be seen in Fig.\,\ref{fig:dos}(i).
 
Finally we consider a zinc vacancy near a Gd atom, the
${\rm Gd_{Zn} +  V_{Zn}}$ defect complex. Its band structure is shown in
Fig.\,\ref{fig:dos}(e).  Extra states appear close to the top of the
valence band. The introduction of a second Gd atom to have a
${\rm Gd_{Zn} + V_{Zn}}$ defect shifts the conduction band, as it can
be seen in Fig.\,\ref{fig:dos}(i) but with no overlap of the Gd states
with other atoms.

\section{Conclusions}

We have investigated ZnO doped with Gd in the presence and absence of
intrinsic defects using density-functional theory and the GW
method.  We show that even the presence of intrinsic
defects can change the distance between the Gd $f$ spin-up and spin-down states. However, this energy difference is not enough to promote a
significant overlap with defect states, which would lead
ferromagnetism in diluted ZnO samples.

\section{Acknowledgement}
The authors are thankful to the Deutsche Forschungsgemeinschaft (DFG) for a Mercator Fellowship under the program FOR1616. A. L. Rosa would like to thanks Sebastien Leb\'egue for fruitful discussions.

\bibliographystyle{apsrev} 
\bibliography{references}

\end{document}